\documentclass[manuscript,screen]{acmart}
\AtBeginDocument{%
  \providecommand\BibTeX{{%
    \normalfont B\kern-0.5em{\scshape i\kern-0.25em b}\kern-0.8em\TeX}}}
\setcopyright{acmcopyright}
\copyrightyear{2022}
\acmYear{2022}
\acmConference[ICAIF '22]{2022 ACM International Conference on Artificial Intelligence in Finance}{November 2022}{New York, USA}
\acmBooktitle{Benchmarks for AI in Finance Workshop} 
\usepackage{microtype}
\usepackage{graphicx}
\usepackage{subfigure}
\usepackage{booktabs} 
\usepackage{float}
\usepackage{hyperref}

\usepackage{tikz}
\usepackage{tikz-network}
\usetikzlibrary{shapes.geometric, arrows}
\tikzstyle{startstop} = [rectangle, rounded corners, minimum width=1.5cm, minimum height=0.75cm,text centered, draw=black, fill=red!30]
\tikzstyle{process} = [rectangle, rounded corners, minimum width=1.5cm, minimum height=0.75cm, text centered, draw=black, fill=orange!30]
\tikzstyle{decision} = [rectangle, rounded corners, minimum width=1.5cm, minimum height=0.75cm, text centered, draw=black, fill=green!30]
\tikzstyle{arrow} = [thick,->,>=stealth]
\tikzstyle{io} = [trapezium, trapezium left angle=70, trapezium right angle=110, minimum width=1.5cm, minimum height=0.75cm, text centered, draw=black, fill=blue!30]
\begin{document}

\title{Towards Evology: a Market Ecology Agent-Based Model of US Equity Mutual Funds}

\author{Aymeric Vie}
\email{vie@maths.ox.ac.uk}
\orcid{0000-0002-7178-1380}
\affiliation{%
  \institution{Mathematical Institute, University of Oxford}
  \country{UK}
}
\affiliation{%
  \institution{Institute for New Economic Thinking at the Oxford Martin School, University of Oxford}
  \country{UK}
}

\author{Maarten Scholl}
\affiliation{%
  \institution{Department of Computer Science, University of Oxford}
  \country{UK}}
\affiliation{%
  \institution{Institute for New Economic Thinking at the Oxford Martin School, University of Oxford}
  \country{UK}
}
\author{Alissa M. Kleinnijenhuis}
\affiliation{%
  \institution{Stanford University}
  \country{USA}
}
\affiliation{%
  \institution{Institute for New Economic Thinking at the Oxford Martin School, University of Oxford}
  \country{UK}
}

\author{James D. Farmer}
\affiliation{%
  \institution{Mathematical Institute, University of Oxford}
  \country{UK}
}
\affiliation{%
  \institution{Institute for New Economic Thinking at the Oxford Martin School, University of Oxford}
  \country{UK}
}
\affiliation{%
  \institution{Santa Fe Institute}
  \country{USA}
}

\renewcommand{\shortauthors}{Vie et al.}

\begin{abstract}
The profitability of various investment styles in investment funds depends on macroeconomic conditions. Market ecology, which views financial markets as ecosystems of diverse, interacting and evolving trading strategies, has shown that endogenous interactions between strategies determine market behaviour and styles' performance. We present Evology: a heterogeneous, empirically calibrated multi-agent market ecology agent-based model to quantify endogenous interactions between US equity mutual funds, particularly Value and Growth investment styles. We outline the model design, validation and calibration approach and its potential for optimising investment strategies using machine learning algorithms.
\end{abstract}


\maketitle

\section{Introduction}
\label{introduction}
\subsection{Motivation}

\paragraph{Research question} One key financial topic of discussion is the Value vs Growth debate. Does Value, focusing on stocks trading for less than their intrinsic or book value, beat Growth, investing in fast-growing companies, over the long term? What investment styles can profit-driven machine learning search find in a multi-fund simulation?

\paragraph{Motivation} A first piece of the answer is undoubtedly macroeconomic and monetary conditions. The last ten years in financial markets have seen the Growth investment style significantly outperform Value. One common reason to justify this performance is the profitability of Growth investing in low-interest rates environments. Growing companies rely on borrowing to fuel their expansion. Hence low-interest rates facilitate this growth. With the recent rising interest rates, and the 2022 bear market particularly affecting tech stocks, those growth prospects appear less favourable and suggest a comeback of the Value style. Outside of those external factors, does Value and Growth performance depend on more endogenous factors such as their respective shares of invested wealth and market composition? Can Value/Growth returns cycles emerge from the endogenous interaction of those different investment styles? Over the last 30 years, this rotation of winners visible in Figure \ref{CIM} in the appendix resembles the oscillations and cycles typically observed in population dynamics. The theory of market ecology \cite{farmer2002market, lebaron2002building, musciotto2018long, lo2019adaptive, levin2021introduction, scholl2021market} borrows concepts from ecology and biology to study financial markets. Trading strategies are analogous to biological species: they exploit market inefficiencies and compete for survival or profit. \cite{scholl2021market} has highlighted the nature of interactions between common trading strategies and the strong density dependence of their returns for stylised trading styles. We here propose to expand this agent-based approach more quantitatively to investigate the Value/Growth interactions in investment funds. We focus on mutual funds: investment companies that pool money from shareholders and invest in securities portfolios.

\paragraph{Related work} This research is in the continuity of the rich area of financial agent-based models and market selection with heterogeneous beliefs \cite{blume1992evolution, blume2006if}. For example, several ABMs have recently been introduced for market-making optimisation \cite{spooner2018market}, understanding flash crashes \cite{paulin2019understanding} and providing sophisticated financial architectures for trading training \cite{byrd2019abides}. We attempt to develop the complementary approach of market ecology \cite{farmer2002market, scholl2021market} by focusing on the ecological interactions between the different types of agents and strategies. 

\paragraph{Significance} This research topic participates in an active area of debate with a novel approach. We describe how some particular results of the market ecology model provide a new, exciting challenge for optimising investment styles using machine learning algorithms. Such simulation-based training can account for interactions and density dependence effects that could be significant and overlooked by traditional time-series training. However, its importance is not limited to the world of financial investment professionals. In the US alone, according to the Investment Company Institute, more than 102 million individuals and an estimated 48\% of households own mutual fund shares \cite{ici22}. The total retirement market assets in the US represent 39 trillion dollars, of which more than 12 trillion are invested in mutual funds. Net sales of regulated open-end funds surged in 2021, with investors placing more than 3 billion dollars in the sector, which holds an increasing share of worldwide equity and debt securities (27\%). Any currently unknown endogenous dynamics at play within the market fund ecosystem thus carry actual, high-magnitude economic impact. 

\subsection{Stylised facts of US equity mutual funds}

Developing a more quantitative agent-based model of the mutual fund industry requires laying down the characteristics of the system we are trying to model. One of the critical elements of validation of the model is its correspondence to the key attributes of real regulated funds. 
At year-end 2021, more than 34 trillion US\$ were invested in open-end funds, which issue new shares and redeem existing shares on demand. This broad category with total net assets of 34 trillion dollars includes mutual funds (27 trillion) and exchange-traded funds (ETFs, 7.2 trillion) but also unit investment trusts (95 billion) and closed-end funds (309 billion). The US contain more than 8,800 mutual funds and 2,800 ETFs. The Investment Company Institute 2022 Fact Book \cite{ici22} describes the total aggregate assets under management and the number of funds of each type. We discuss modelling approaches (few aggregate agents vs many agents) and present some mutual fund data in the appendix.

\section{Model}
\label{model}
We consider a population of $n$ investment funds, which trade shares of a single representative asset and cash. Every period-day $t$, the funds can buy, sell and short-sell shares of the asset in constant supply. Asset shares pay daily dividends $\delta(t)$ following an autocorrelated Geometric Brownian Motion. Cash yields an interest rate $r$ paid daily. We build over the model of \cite{scholl2021market}, with notable improvements in empirical calibration. We show an example simulation run in the appendix.


\begin{figure}[ht]
\centering
\begin{tikzpicture}[node distance=1cm]
\node (start) [startstop] {Initialise};
\node (pro2) [process, right of=start, xshift=1.5cm] {Trading signal};
\node (pro3) [process, right of=pro2, xshift=2cm] {Demand};
\node (pro4) [process, below of=pro3, yshift=-0.5cm] {Market clearing};
\node (pro5) [process, below of=pro4, yshift=-0.5cm] {Dividends/interest};
\node (pro7) [process, below of=pro2, yshift=-0.5cm] {Solvency};
\node (pro6) [process, below of=pro2, yshift=-2cm] {Investment};
\node (stop) [startstop, below of=start, yshift=-2cm] {Terminate};
\draw [arrow] (start) -- (pro2);
\draw [arrow] (pro2) -- (pro3);
\draw [arrow] (pro3) -- (pro4);
\draw [arrow] (pro4) -- (pro5);
\draw [arrow] (pro5) -- (pro6);
\draw [arrow] (pro7) -- (pro2);
\draw [arrow] (pro6) -- node[anchor=west] {For $T_{\text{max}}$ periods} (pro7);
\draw [arrow] (pro6) -- node[anchor=north, yshift = -0.35cm] {After $T_{\text{max}}$ periods} (stop);
\end{tikzpicture}
\caption{Visual summary of the model components. }
\label{model_summary}
\end{figure}
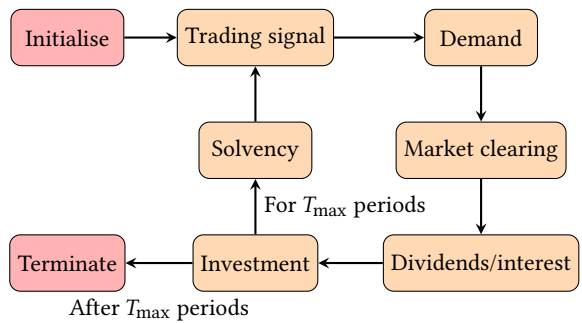

\subsection{Trading strategies and signals}
Like real markets, our financial market model features a diverse sample of stylised versions of the most common funds' trading strategies \cite{scholl2021market}. The previous section outlined the main results from the ICI data \cite{ici22}, which suggests including Value and Growth funds, possibly divided by their cap, in the model. For the current version of the model in development, we include three styles: Value, Noise trading and Momentum (trend following). \textit{Value investors} (VI) form heterogeneous subjective valuations $V_i$ of the asset based on discounted sums of dividends. \textit{Noise traders} (NT) trade on a similar valuation perturbed by a mean-reverting Ornstein Uhlenbeck process $X(t)$, mimicking exogenous sentiment dynamics. We calibrate the parameters of the Ornstein Uhlenbeck process to match empirical excess volatility \cite{scholl2021market}. \textit{Trend followers} (TF) trade on the existence of trends in the asset price over various time horizons. Agents' trading strategies are represented by their \textit{trading signals} $\phi(t)$.

\begin{align}
    \phi_i^{\text{NT}}(t) = \log_2\left(X_i(t)V_i(t) / p(t)\right)   \\
    \phi_i^{\text{VI}}(t) = \log_2\left(V_i(t) / p(t)\right) \\
    \phi_i^{\text{TF}}(t) = \log_2\left(p(t-1) / p(t-\theta_i)\right)
\end{align}

\subsection{Asset demand}

The funds' daily trading signals are inputs of the \textit{excess demand function} for the asset \cite{scholl2021market}. The excess demand function expresses the demand of the fund for the asset as a function of the unknown price $p(t)$. Fund wealth $W$ is the sum of agent cash, present value of asset shares and liabilities. Our demand function features maximum leverage ${\lambda}$ and strategy aggression $\beta$. Our demand function simply represents an investor with asset position $S(t)$ and budget $\lambda W(t)$ spending a share $\Tilde{\phi}(t)$ of her budget on the asset, and $1-\Tilde{\phi}(t)$ on the cash \cite{poledna2014leverage, scholl2021market}.

\begin{equation}
    D(t, p(t)) =  \Tilde{\phi}(t) \frac{\lambda W(t)}{p(t)} - S(t) 
\end{equation}

For $\Tilde{\phi}(t) = \tanh (\beta \phi(t))$: the $\tanh$ function smooths and bounds the trading signal in the range $[-1,1]$, so that the demand never exceeds the agent budget including leverage. This demand function is continuous, allows short-selling, and enforces deleveraging \& margin calls\footnote{Previous designs of the demand function in our model tended to generate huge short positions. Because of the embedded deleveraging, i.e. reduction of short positions in case of price increase, this demand function over the simulation test run gives an average short position size equivalent to $1.03\%$ of asset supply. This level is in line with the top10 NASDAQ stocks ($1.17\%$ of float on average). If leverage increases, the short ratio increases to riskier stock levels.}, and always deliver orders that respect the budget constraint.

\subsection{Market-clearing}

The \textit{market-clearing} process finds the price for which the sum of the funds' demands equals the fixed asset supply $Q$, demand matching supply \cite{poledna2014leverage}. This is equivalent to the market-clearing procedure of finding the root of the aggregate excess demand function \cite{scholl2021market}. The market-clearing condition here is thus: $\sum_iD_i(t, p(t)) = Q$. While many financial agent-based models use limit-order books (LOBs), our focus here is on long timescales from decades to centuries, while those models focus on intraday dynamics. We do not exclude using LOBs in this model's future but believe that market clearing is sufficient for our time horizon.

\subsection{Dividends, interest and investment flows}

After computing the clearing price, funds execute the resulting demand orders. Agents receive capital gains: the dividends $\delta(t)$ and interest $r$ corresponding to their new positions. The actions of external investors play an essential role in the wealth dynamics of mutual funds. Depending on the performance of the funds, external investors can choose to redeem their shares or buy new fund shares. Our model models those inflows and outflows in the investment module according to empirical data on fund flows.

\subsection{Solvency}

Funds with negative wealth enter bankruptcy and exit the market. An administrator slowly liquidates their shares. The wealthiest fund will split into several identical, equal-sized entities to fill the vacant spot. This mechanism keeps the number of funds and asset shares constant and limits market perturbations due to insolvencies.

\section{Calibration and Validation}
\label{calibration}
Calibration and validation of agent-based models (ABMs) are crucial \cite{paulin2018agent}. A common criticism of ABMs is that they often have too many parameters and risk being unrealistic. Our validation includes three main targets. The first is for the model to reproduce the stylised financial properties of asset returns \cite{cont2001empirical, cont2007volatility}. The second is to model realistic fund flows. The third is for the fund agents to be consistent in various properties with the empirical data on mutual funds \cite{ici22}: their number/size, returns, and investment styles. While we achieve the first two, more work is necessary to satisfy the latter, as we detail in the appendix. 



\subsection{Reproducing stylised facts of financial markets}

Generating the so-called financial ``stylised facts'' is a popular requirement for validating financial ABMs. Our model reproduces the main stylised facts of asset prices \cite{cont2001empirical}. Our log prices display intermittency. The log price returns do not show significant linear autocorrelations past trivial frequencies. Returns show heavy tail distributions with excess kurtosis compared to a normal distribution. We can also reproduce the leverage effect -negative correlation between price returns and volatility- a positive volume-volatility correlation and slow decay of autocorrelation in absolute returns. We provide more details in the appendix.

\subsection{Calibrating Investment flows}

We calibrate model investment flows to ensure that funds' returns lead to realistic sales and redemption flows, as these flows represent a significant share of funds' net value. Investment companies are subject to reporting requirements of their assets, redemptions, sales, and other indicators through various SEC forms. \cite{ha2019misspecifications} analysed  N-SAR reports and fund returns and identified a linear, positive relationship between funds' excess return and investment flow, suggesting that external investors are chasing returns, confirming earlier results \cite{chevalier1997risk}. We estimated linear regression models to predict net fund flows from fund excess returns at various lags. Indeed, investors may look at fund returns on a monthly, yearly or even multi-year basis. We identified a few significant predictors of net flows. The constant regression term is negative, illustrating investors redeeming their shares for liquidity purposes. 10-year excess returns have a positive coefficient, and the constant is negative. 

\section{Results}
\label{results}
We present some vanilla dynamics of the Evology ABM without strategy evolution: the asymptotic distribution of wealth between strategies. We are interested in their dependency on the initial wealth distribution. On each point of a uniform sample of $500$ points in the three-dimensional simplex, we run the simulation $10$ times for $150$ years of trading. We measure the average wealth share of the last $10,000$ days of trading. This sampling ensures convergence in wealth distributions and accounts for stochasticity. The returns and wealth shares of the strategies significantly vary with their position in the simplex, showing density-dependence \cite{farmer2002market, scholl2021market}. Experiment parameters ensure that we observe the asymptotic wealth shares after convergence. Figure~2 provides an example of the price and dividend series generated during a single simulation run, and appendix Table~2 presents the performance of the base strategies in the example run. \par 

Figures~3-5 show the final wealth shares of each strategy after 150 years, depending on the initial condition. Outside a specific corner of the simplex, the wealth percentage of noise traders goes down to negligible levels of 5 to 20\%. Value investors are almost absent from the left boundary but dominate most of the simplex configurations with a majority of 70\% of the wealth. Trend followers dominate the unstable top region and are absent around the bottom axis. Early extinction of the other strategies characterises this unstable top region: if initialised in high proportion to other strategies, TFs are detrimental to the other species. \par

\section{Trading Strategy Optimisation in Evology}
\label{discussion}
Trading strategy search is a popular topic for applying machine learning. Quantitative trading systems driven by linear \& logistic regression, support vector machines, reinforcement learning, deep neural networks, random forests, genetic algorithms, and genetic programming have successfully created profitable strategies \cite{allen1999using,  dempster2001real, zhang2016using, ta2018prediction, hasan2020modeling}. Derivative-free methods are relevant since the profit objective usually cannot be reduced to an explicit, derivable objective function. Evolution strategies and program synthesis can also be particularly performant for this task. \par 

The returns of a trading strategy depend on the wealth distribution of the market, i.e. the wealth shares owned by the different strategies \cite{scholl2021market}, challenging optimisation. Strategies experience crowding: their returns decrease as their size grows and exceeds the carrying capacity of their niche \cite{farmer2002market, scholl2021market}. Offline tuning, which assumes no interactions between the strategy and the training data, can overlook this effect. These additional difficulties invite reassessing the performance of those popular machine learning approaches for trading strategy search in a simulation environment. In addition, provided that the simulation environment is realistic enough, training on the data-generating process behind market dynamics can lead to more robust strategies than training on single sample paths of this market process. 

\subsection{Benchmark learning tasks}

\paragraph{General challenges and environments} Beyond the tasks described below, the strategies evolved in Evology should satisfy some higher-end goals. We do not desire the machine learning algorithms to result in incomprehensible, over-fitting strategies to maximise profits. For the trained strategies to be interesting, they need to be \textit{interpretable} -display some level of economic insight- and \textit{robust} -successfully operate under various market conditions-. We develop in more detail those general challenges and provide more details on the tasks mentioned below in the appendix. We can consider two different environments: an environment of \textit{stable strategies} where the strategies of all the agents (except the optimising agent) stay fixed over time. The alternative is a \textit{coevolution} setting where a small fraction of the population is adaptive, imitating the activity of actively managed/hedge funds. This adaptation could use an extensive range of models from the intensity of choice and imitation to more sophisticated evolutionary or machine learning approaches. This coevolution setting brings an additional challenge to strategy optimisation.

\paragraph{Task 1: Trading strategy optimisation} We are interested in an individual \textit{fund} optimising its trading strategy $\phi(t)$ to maximise profits during a $T$-period market run, using a profit measure such as the cumulative return or the Sharpe ratio. The baseline levels to achieve would be i) become more profitable than the base strategies and ii) become more profitable than empirical strategies.

\paragraph{Task 2: Investment strategy optimisation} We are interested in an individual \textit{investor} optimising its investment strategy. The investor is learning how to invest, i.e. a function $\nu$ mapping the various fund characteristics (return, size...) to positive or negative investment amounts, intending to maximise the profitability of their investments.

\section{Conclusion}
\label{conclusion}
We present Evology\footnote{Available open-source \url{https://github.com/aymericvie/evology}}, an empirically calibrated financial agent-based model of US equity mutual funds grounded on the market ecology perspective. The complexity of strategies' interactions and the density-dependence of returns make this specific optimisation problem challenging for search algorithms: dynamic, deceptive, and no clear optimal solution. This multi-agent model has the potential to become a new training ground for investment strategy search with machine learning methods. We further discuss the model limitations and next steps in the appendix.


\begin{acks}
This publication is based on work supported by the EPSRC Centre for Doctoral Training in Mathematics of Random Systems: Analysis, Modelling and Simulation (EP/S023925/1) and by Baillie Gifford. The authors thank Christian Schroeder de Witt for helpful advice, Yeonjeong Ha and Kwangsoo Ko for sharing their N-SAR and CRSP data \cite{ha2019misspecifications}, the Mathematical Institute of the University of Oxford for the use of computing servers, the ICML22 AI4ABM workshop where a previous version of this research was discussed \cite{vie2022evology}, members of Oxford INET and Fidelity Investments for advice and anonymous referees for their valuable suggestions.
\end{acks}

\bibliographystyle{ACM-Reference-Format}
\bibliography{main}

\appendix
\section*{Appendix}


\subsection{Value and Growth performance}

\begin{figure}
    \centering
    \includegraphics[width=0.5\textwidth] {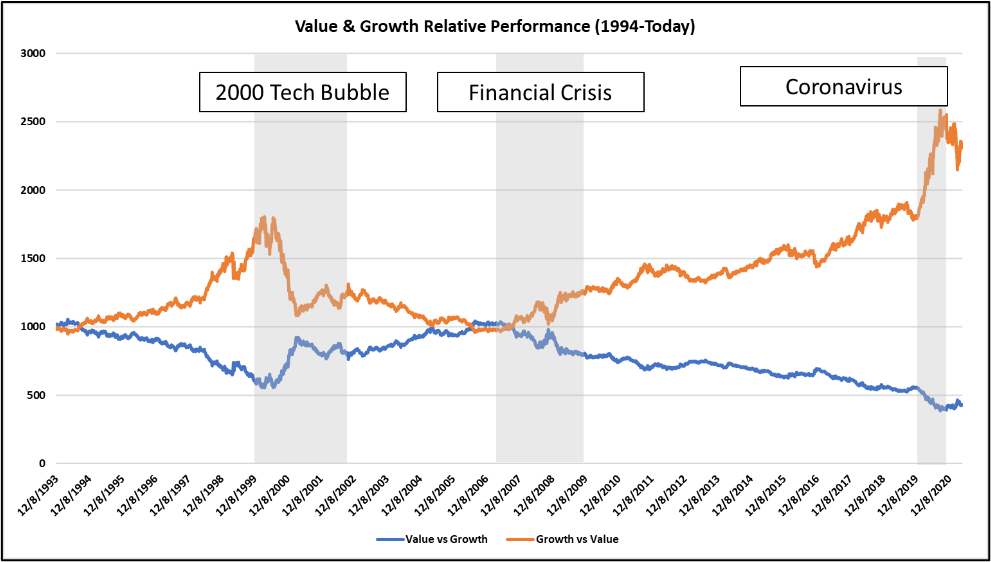}
    \caption{Value and Growth relative performance (1994-2021). Source: Canterbury Investment Management, "Value versus growth: a brief historical view"}
    \label{CIM}
\end{figure}

\subsection{Mutual Funds: overview of ICI data}

\begin{table}
  \caption{Key elements on Equity mutual funds' investment styles at year-end 2021 \cite{ici22}}
  \label{facts}
  \begin{tabular}{ccc}
    \toprule
    Investment styles & TNA (million US\$) & Number of funds\\
    \midrule
    Growth small-cap & 315,759 & 168 \\
    Growth mid-cap & 354,192 & 164 \\
    Growth large-cap & 1,131,329 & 247 \\
    Growth multi-cap & 991,526 & 140 \\
    \midrule
    Value small cap & 197,397 & 194 \\
    Value mid cap & 269,347 & 172 \\
    Value large cap & 811,646 & 279 \\
    Value multi-cap & 377,509 & 167 \\
    \midrule
    Blend small-cap & 338,924 & 179 \\
    Blend mid-cap & 473,702 & 133 \\
    Blend large-cap & 3,741,454 & 380 \\
    Blend multi-cap & 1,679,108 & 221 \\
  \bottomrule
\end{tabular}
\end{table}

\subsection{Model validation - Empirical stylised facts of asset prices \cite{cont2001empirical}}

The following figures related to stylised facts of asset prices use data from a single simulation run, starting from the $[1/3, 1/3, 1/3]$ initial condition on wealth shares. The interest rate equals $0.01$, external investment is inactive, and the random seed is equal to $0$ for reproducibility of the dividend and noise processes.

\paragraph{Absence of autocorrelations} Autocorrelations of asset returns should be insignificant except for very small time scales, in which the microstructure has some impact.

\begin{figure}[H]
    \centering
    \includegraphics[width=0.5\textwidth]{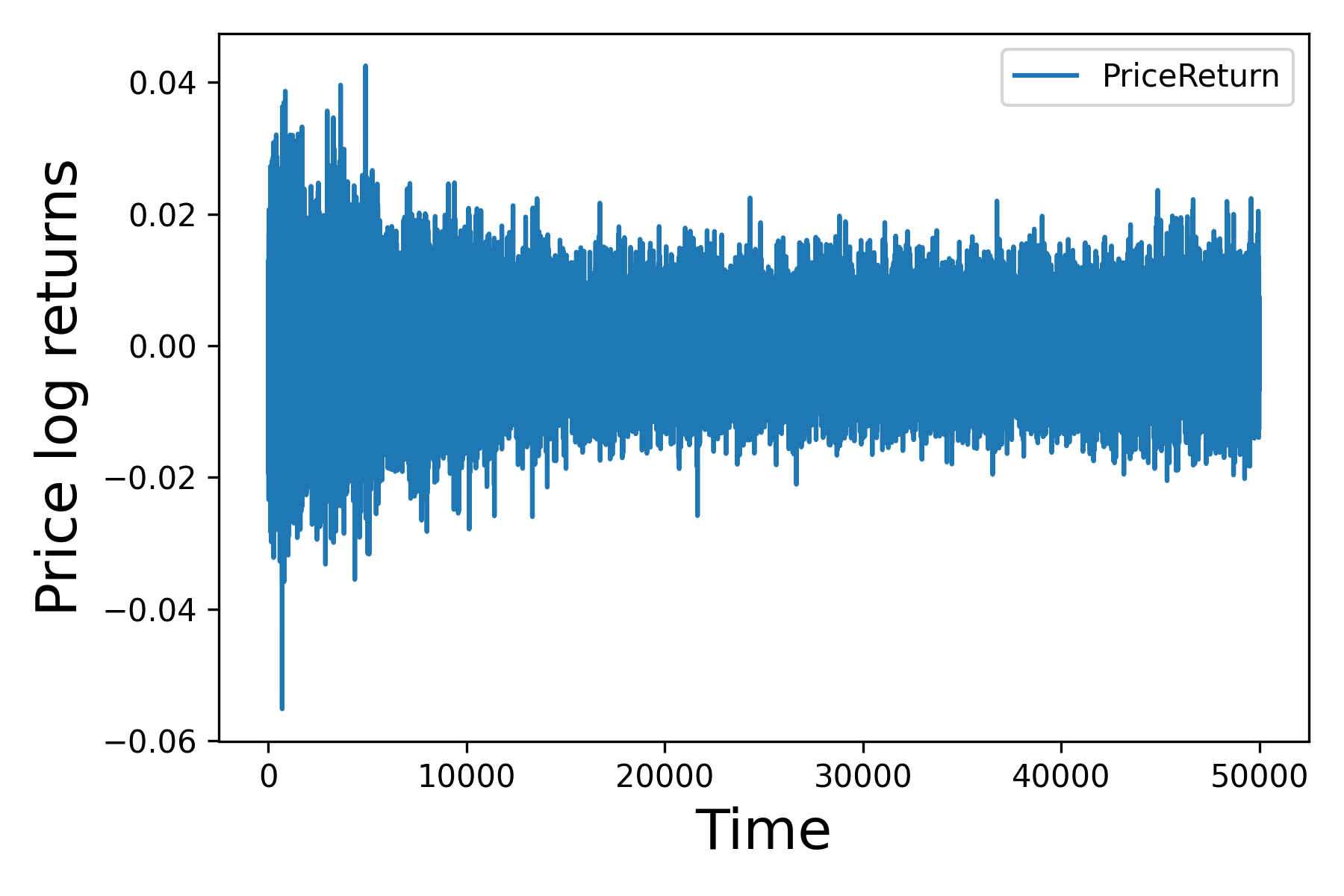}
    \caption{Daily log returns of the asset price. For a price $p(t)$, the log return at the daily timescale is $r(t) = \ln{p(t)} - \ln{p(t-1)}$.}
    \label{log_returns}
\end{figure}

\begin{figure}[H]
    \centering
    \includegraphics[width=0.5\textwidth]{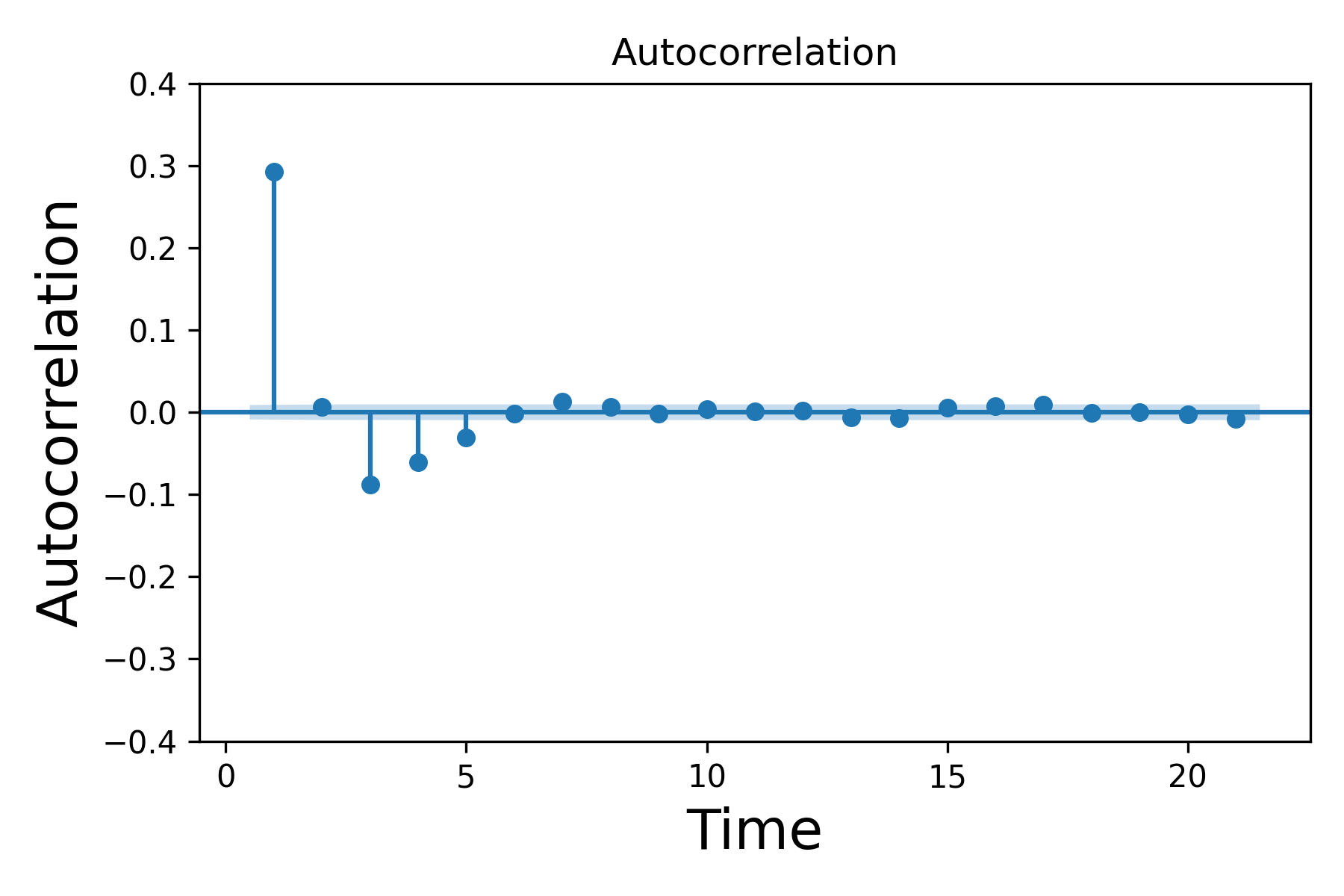}
    \caption{Autocorrelation function of log returns of the asset price for up to 21 periods. After a short timescale of 5 periods, autocorrelations are not significant from zero, which confirms the absence of autocorrelations.}
    \label{log_returns_acf}
\end{figure}

\paragraph{Heavy tails} The unconditional distribution of returns should display a power-law or a Pareto tail, with finite variance, excluding the normal distribution.

\begin{figure}[H]
    \centering
    \includegraphics[width=0.5\textwidth]{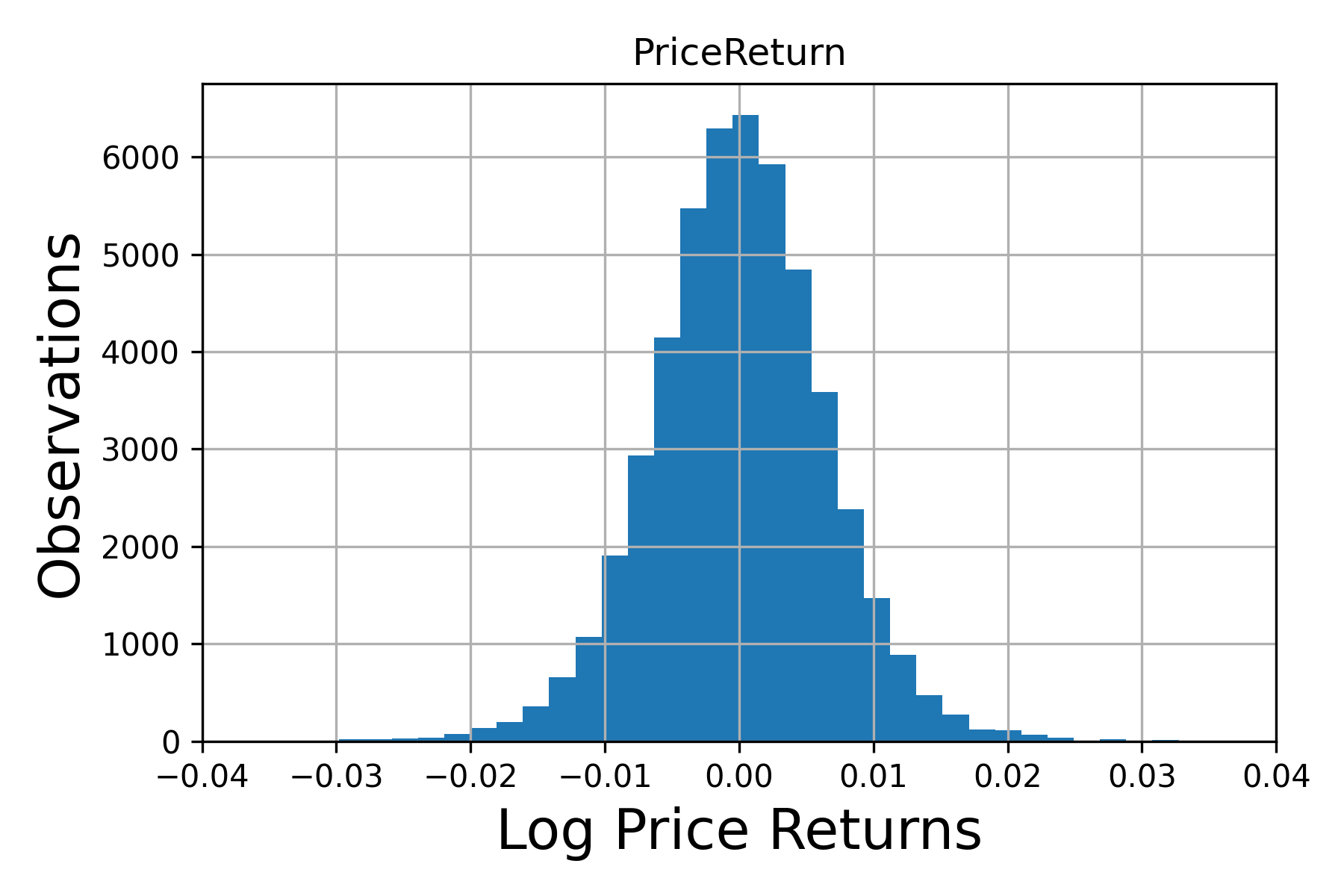}
    \caption{Histogram of the daily log price returns shows heavy tails. The Fisher's (or excess) kurtosis ($\kappa$) value for the series is $1.3$, which is superior to Fisher's kurtosis of the normal distribution, which is equal to $0$. $\kappa = E\left[\left(\frac{X-\mu}{\sigma}\right)^4\right] - 3$.}
    \label{heavy_tail_returns}
\end{figure}

\paragraph{Gain/loss asymmetry} One should observe large drawdowns in stock prices without equally large upward movements. 

\begin{figure}[H]
    \centering
    \includegraphics[width=0.5\textwidth]{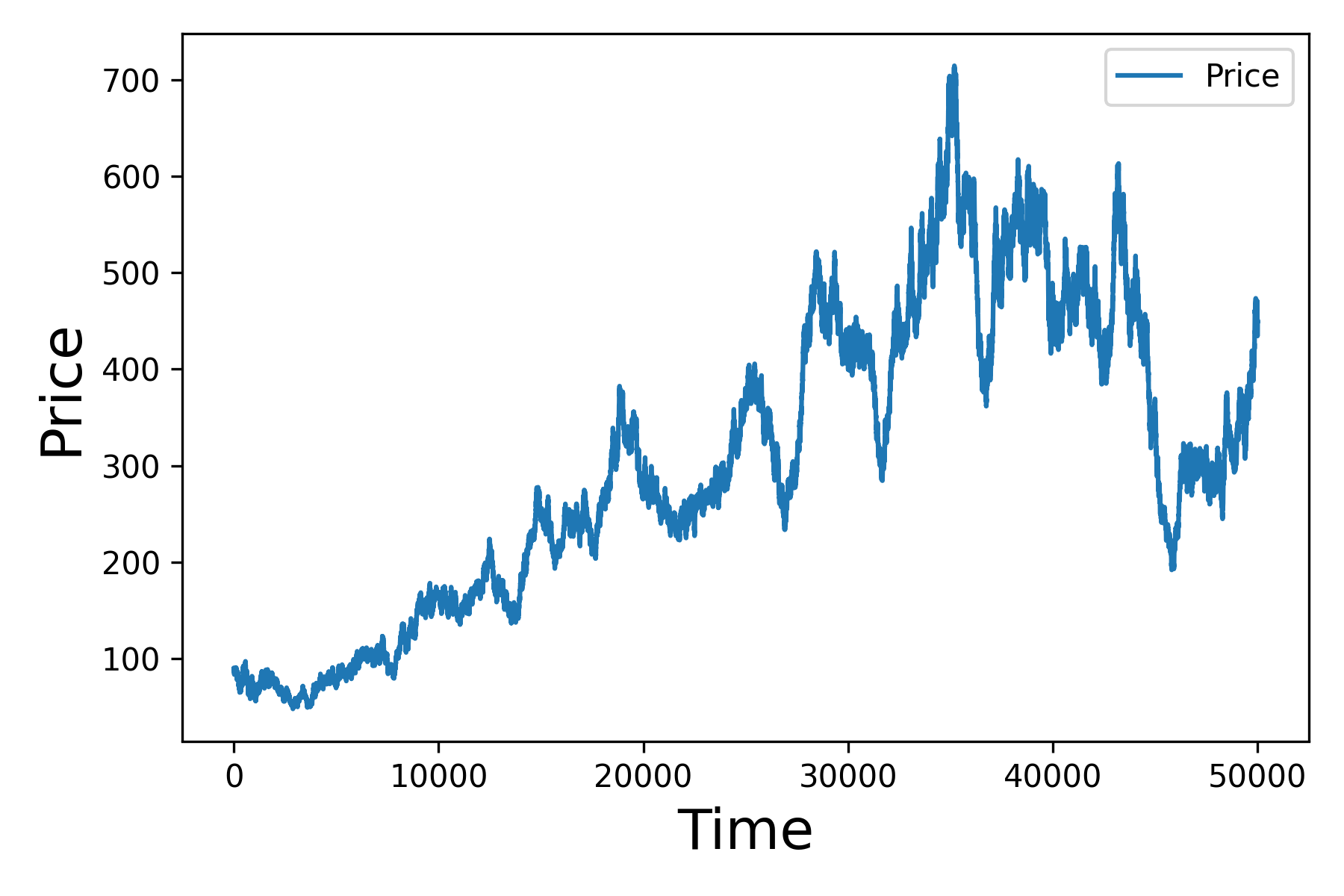}
    \caption{Asset price series of the simulation run. Graphical analysis highlights faster downward than upward price movements: it usually takes more time periods to recover from a drawdown than the drawdown duration.}
    \label{regular_price}
\end{figure}

\paragraph{Aggregational Gaussianity} As we increase the timescale for calculating the returns, their distribution should look more and more like a normal distribution.

\begin{figure}[H]
    \centering
    \includegraphics[width=0.5\textwidth]{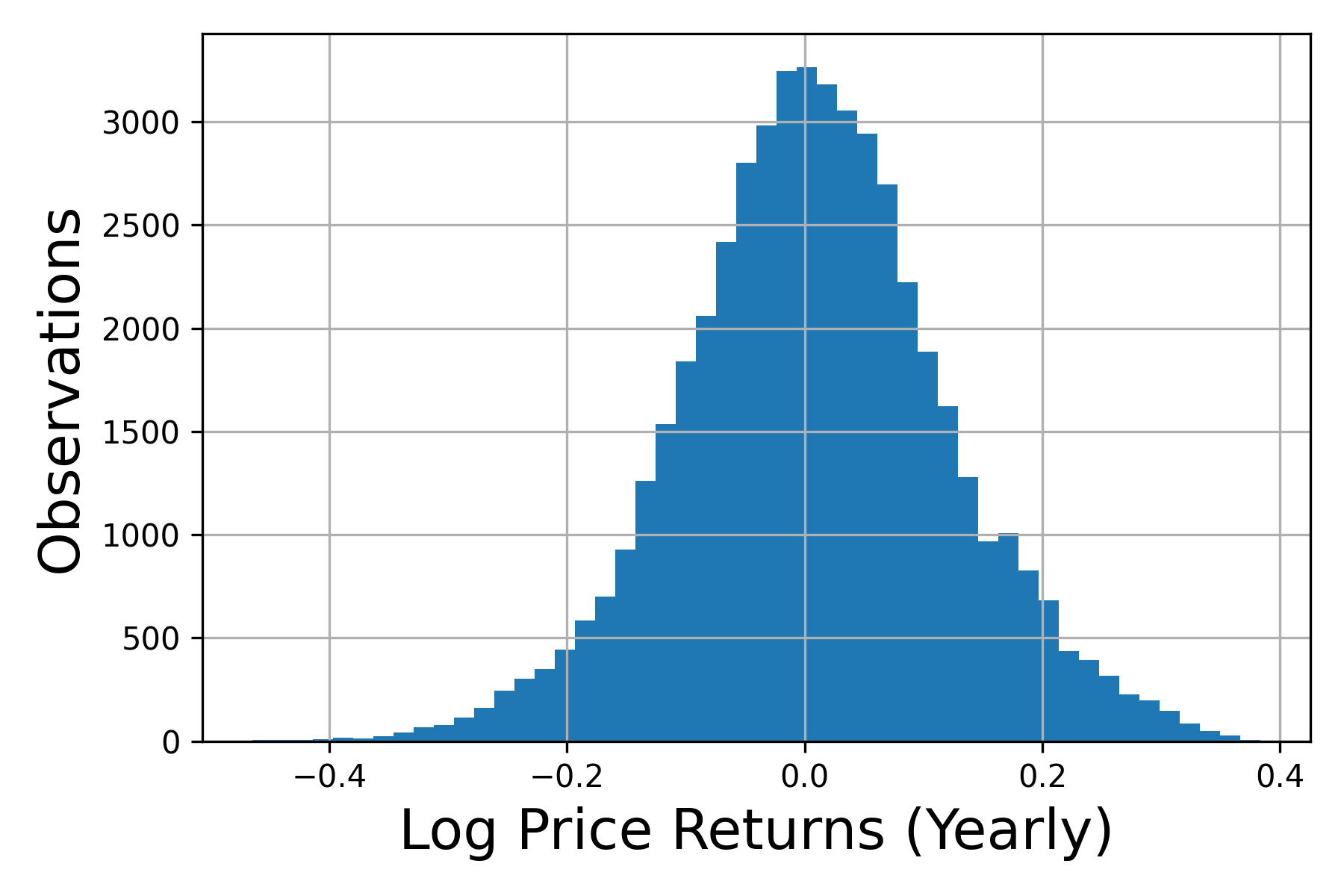}
    \caption{Distribution of yearly log price returns is closer to a normal distribution than the distribution of daily log price returns in Figure \ref{heavy_tail_returns}. Quantitatively, the excess kurtosis for monthly returns is lower, equal to $0.85$, and the excess kurtosis of yearly returns is even lower at $0.32$. Hence, we establish aggregational Gaussianity.}
    \label{distribution_yearly_returns}
\end{figure}

\paragraph{Intermittency} Returns should display high variability, visible by the presence of irregular bursts in time series of volatility estimators.

\begin{figure}[H]
    \centering
    \includegraphics[width=0.5\textwidth]{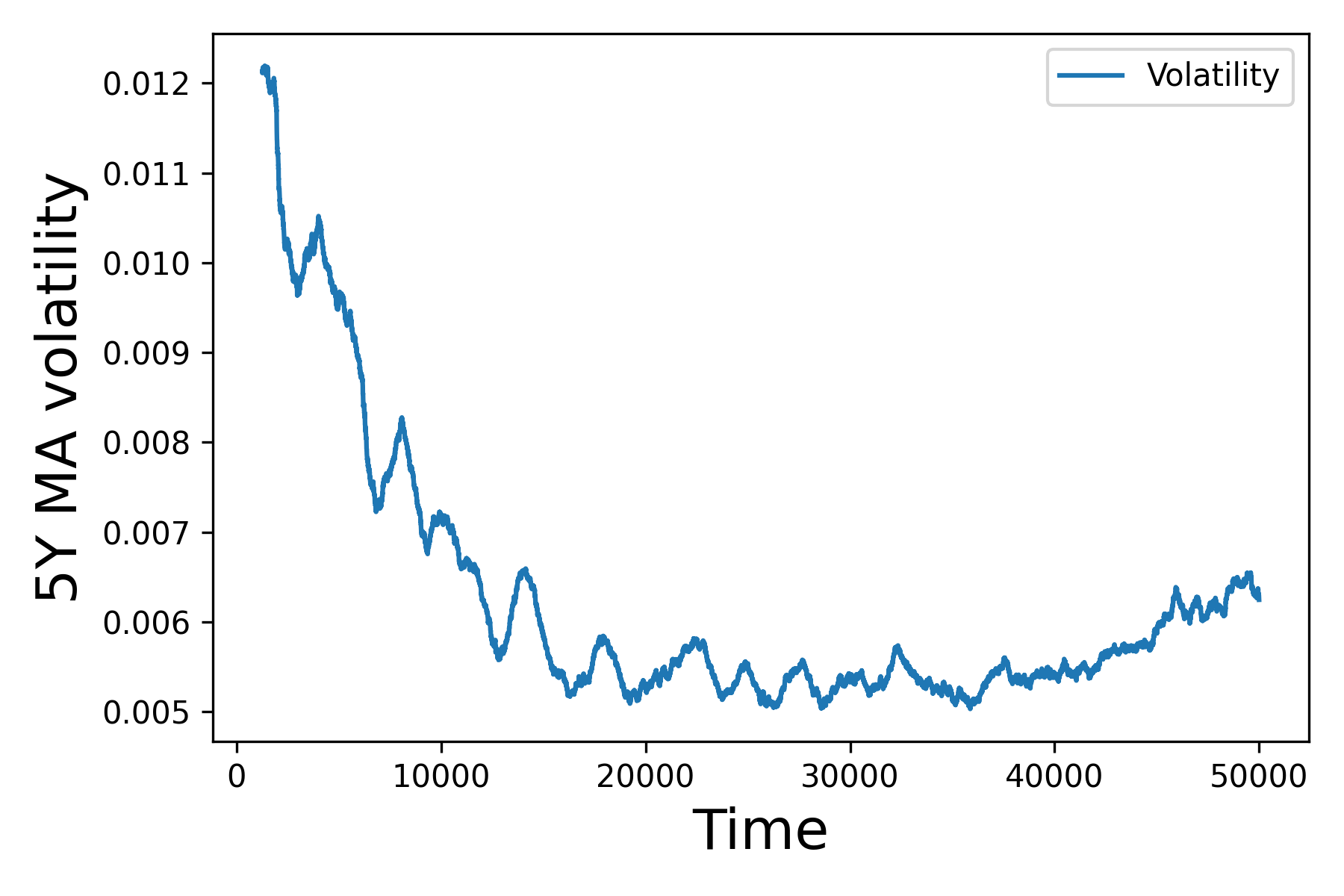}
    \caption{5-year moving average volatility over time show intermittency. We attribute the relatively higher level of volatility in the early steps from the runs to the transient period due to the agent-based model initialisation. However, the irregular bursts in the log-returns of Figure \ref{log_returns} are not obvious, perhaps given the large number of periods represented.}
    \label{intermittency}
\end{figure}

\paragraph{Volatility clustering and slow decay of autocorrelation in absolute returns} Different volatility measures should display a positive autocorrelation over several days, showing that high-volatility events tend to cluster over time. We can observe this volatility clustering from the slow decay of the autocorrelation function of absolute returns.

\begin{figure}[H]
    \centering
    \includegraphics[width=0.5\textwidth]{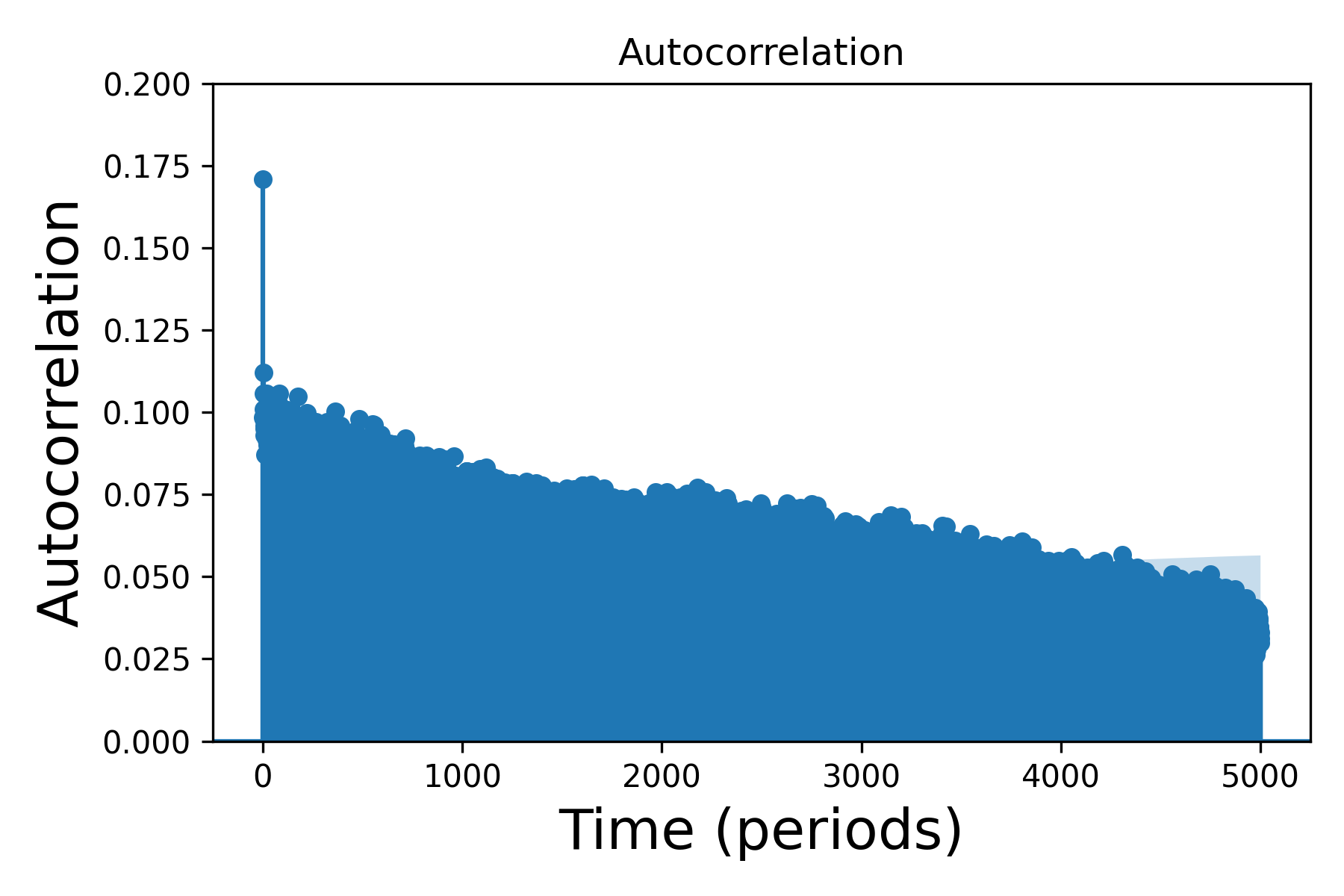}
    \caption{Autocorrelation function of absolute log price returns show long range dependence and slow decay as a function of time. The autocorrelation remains significantly positive over many time lags, showing volatility clustering \cite{cont2007volatility}.}
    \label{volatility_clustering}
\end{figure}

\paragraph{Leverage effect} Measures of asset volatility should negatively correlate with the asset returns. Over our simulation run, this Pearson correlation is negative and equal to $-0.005$.

\paragraph{Volume/volatility correlation} Trading volume should correlate with volatility measures. Over our simulation run, this Pearson correlation is positive and equal to $0.48$. In addition, volume positively correlates with price returns ($0.001$).

\subsection{Simulation run example}

\begin{figure}[H]
    \centering
    \includegraphics[width=0.5\textwidth]{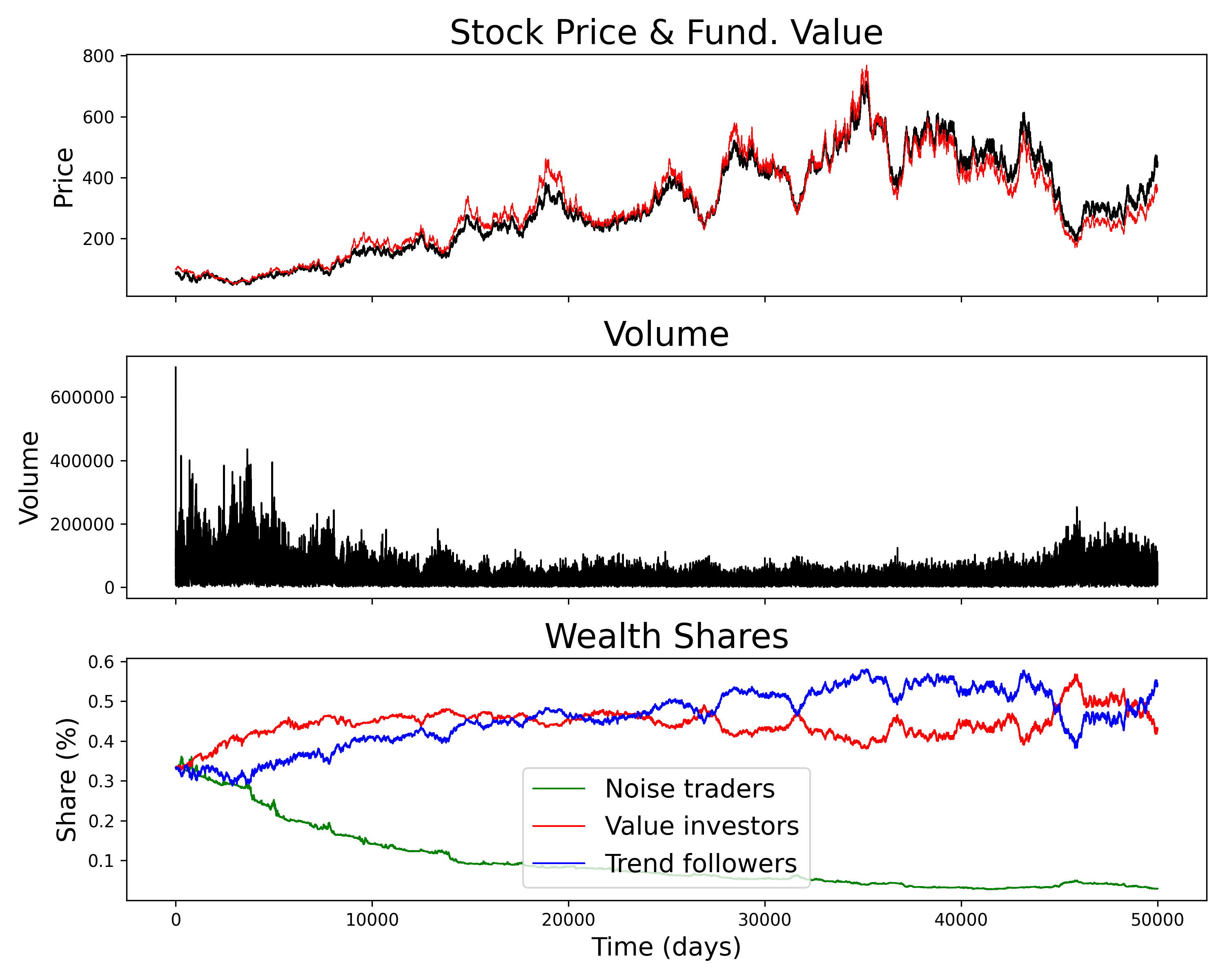}
    \caption{Price, fundamental value, volume and strategy types wealth shares during a 50,000-day (around 200 years) simulation run, starting from the initial coordinates $[1/3, 1/3, 1/3]$.}
    \label{price}
\end{figure}

\begin{figure}[H]
    \centering
    \includegraphics[width=0.5\textwidth]{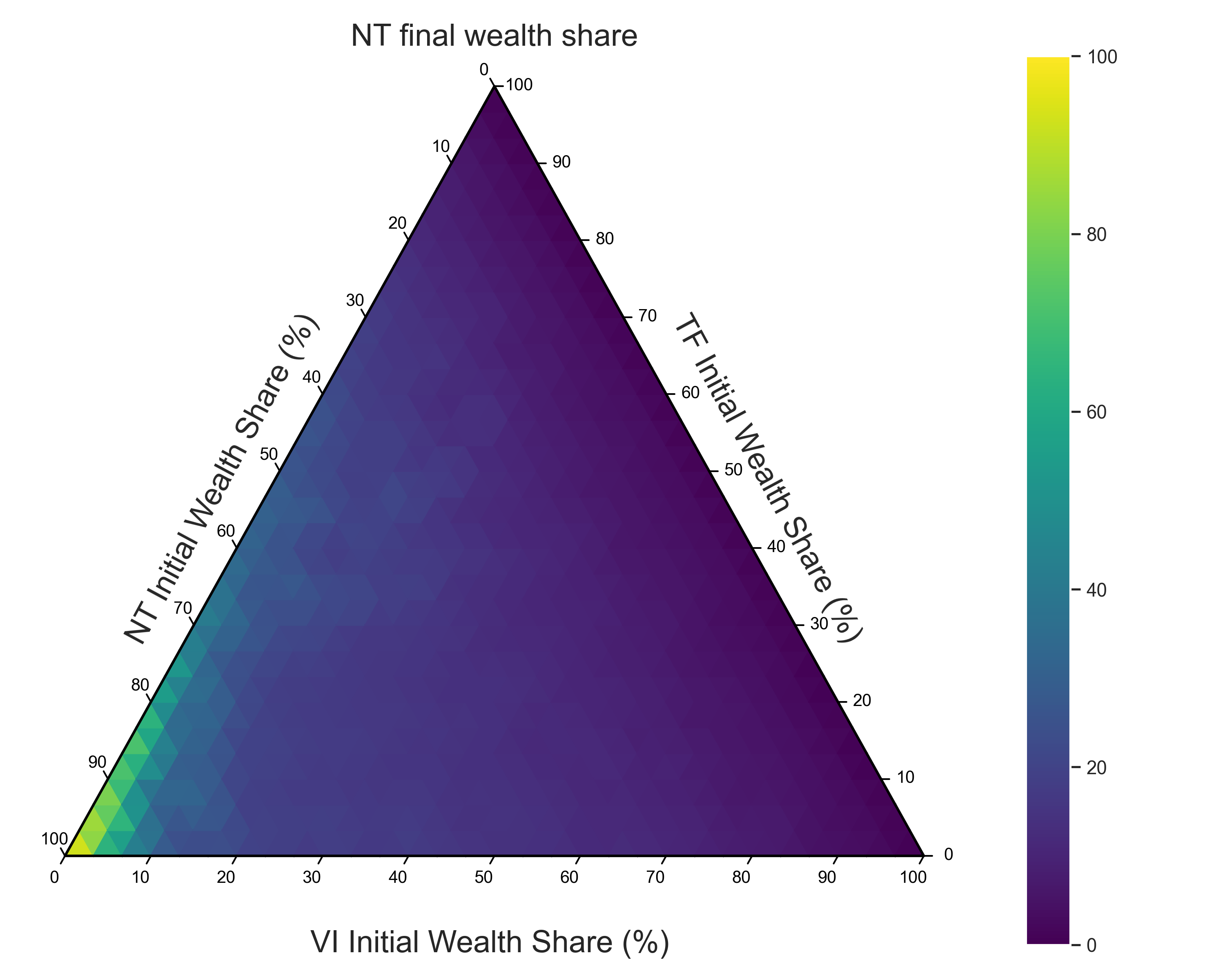}
    \caption{Noise traders' final wealth share}
    \label{ws_nt}
\end{figure}

\begin{figure}[H]
    \centering
    \includegraphics[width=0.5\textwidth]{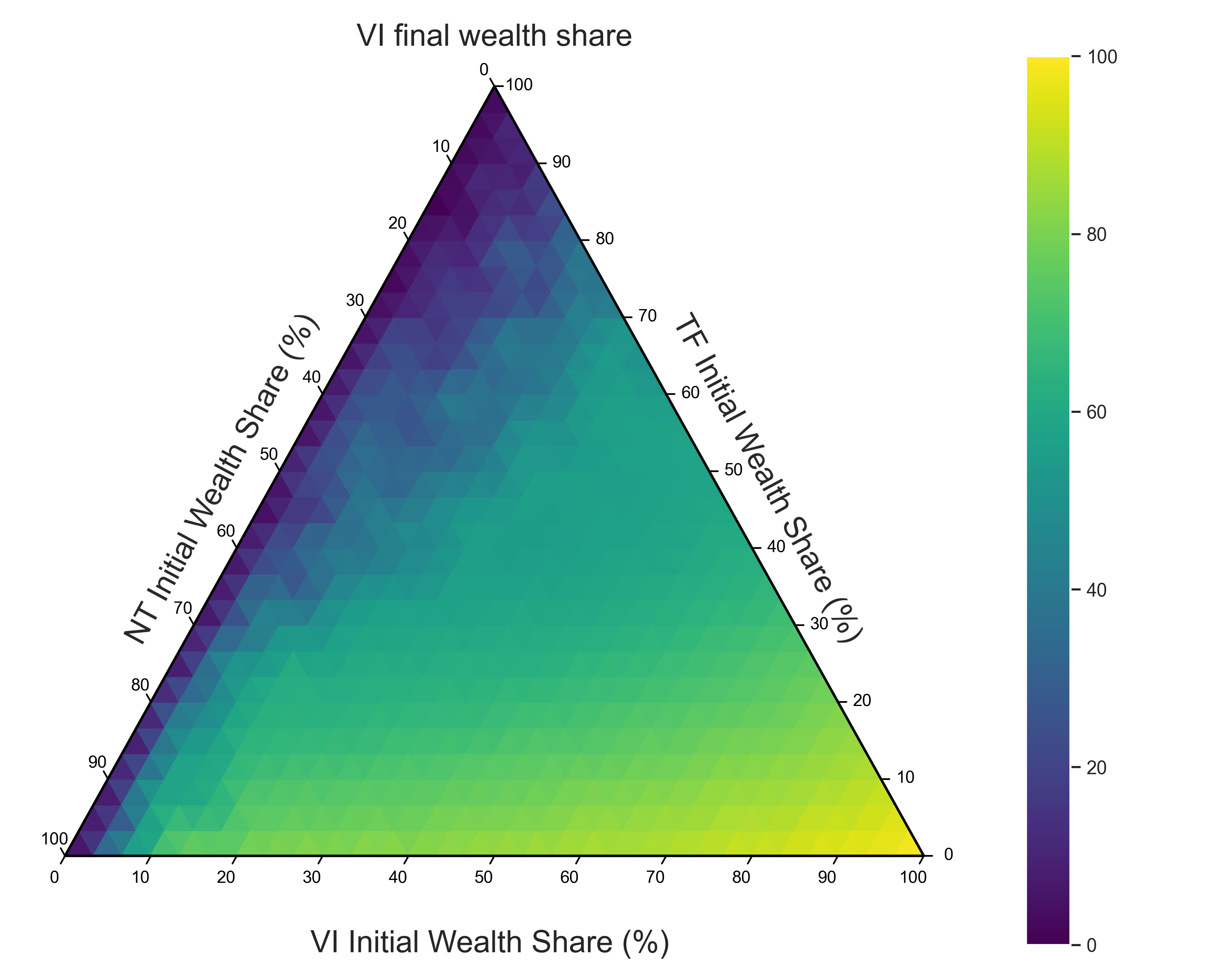}
    \caption{Value investors' final wealth share}
    \label{ws_vi}
\end{figure}

\begin{figure}[H]
    \centering
    \includegraphics[width=0.5\textwidth]{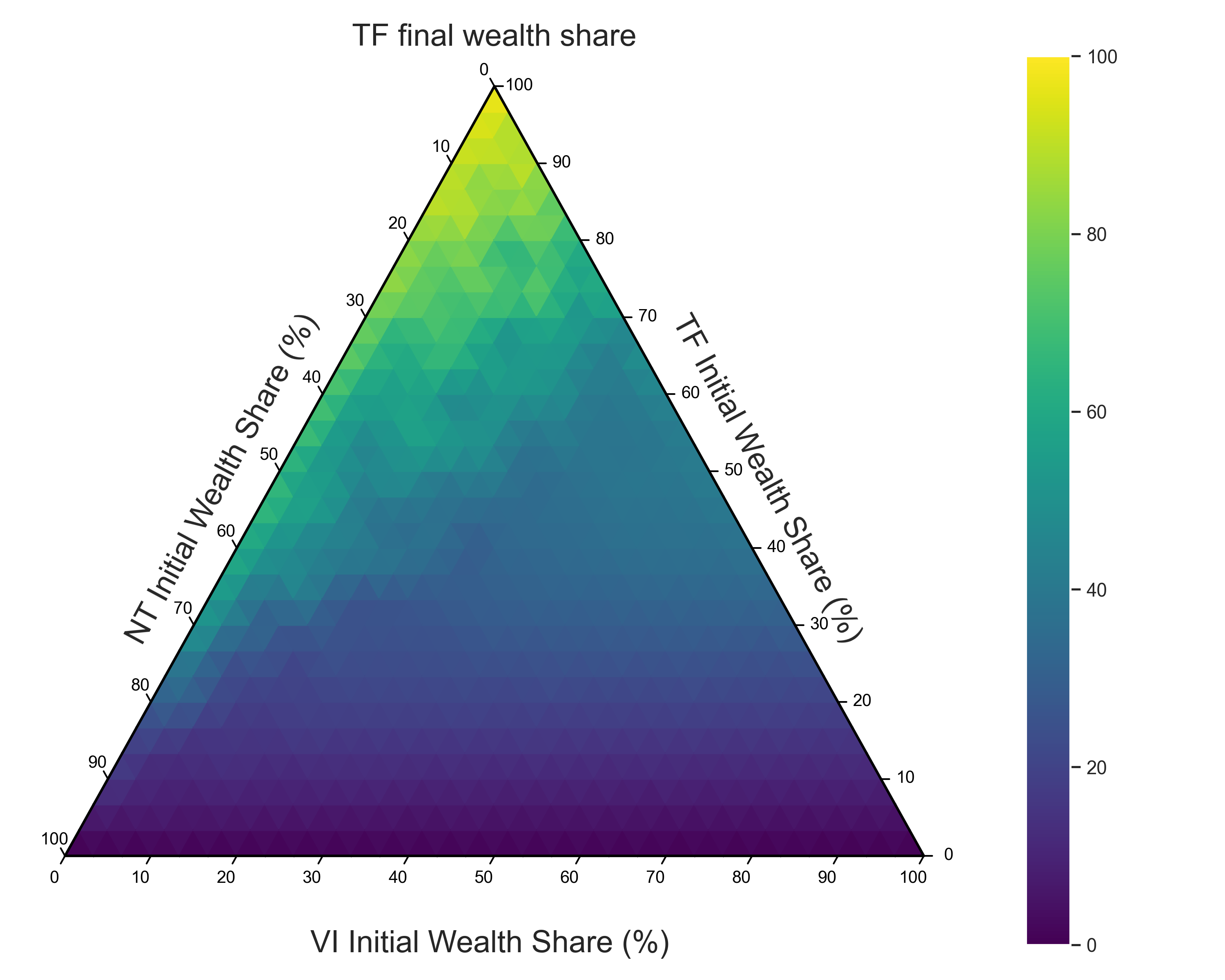}
    \caption{Trend followers' final wealth share}
    \label{ws_tf}
\end{figure}

\begin{table}[t]
\caption{Returns (daily and yearly geometric mean, in \%) and Sharpe ratios (daily and yearly for 252 trading days) of base strategies at the initial coordinates $[1/3, 1/3, 1/3]$. Base strategies are: NT, VI, TF, BH (buy and hold), IR (interest rate). The IR strategy has constant returns hence no standard deviation.}
\label{returns-table}
\vskip 0.15in
\begin{center}
\begin{small}
\begin{sc}
\begin{tabular}{lccccc}
\toprule
& NT & VI & TF & BH & IR \\
\midrule
Returns (d)  & 0.007 & 0.007 & 0.006 & 0.007 & 0.004 \\
Returns (y) & 1.668 & 1.806 & 1.424 & 1.687 & 1.005 \\
Sharpe (d) & 0.029 & 0.026 & 0.02 & 0.14 & NA
\\
Sharpe (y) & 0.463 & 0.41 & 0.317 & 0.222 & NA \\
\bottomrule
\end{tabular}
\end{sc}
\end{small}
\end{center}
\vskip -0.1in
\end{table}

\subsection{Estimating fund performance-flow relationship}

We employ two separate procedures to estimate how the fund performance impacts net cash flows, i.e. inflows minus outflows expressed as a function of the fund TNA. The first uses fund-level data with the SEC Form N-SAR filings 
\cite{ha2019misspecifications}, and the second aggregate data from the Investment Company Institute \cite{ici22}. Both sources confirm that 10-year returns positively influence net flows, with similar results on both data sources. Net flows follow a constant decrease of a few percentage points of TNA, corresponding to the higher-level wealth movements from active to index funds and investor portfolio rebalancing during equity gains.

We define the fund return $r_k(t)$ between period $t$ and period $t-k$, with $f(t)$ estimated net flows at period $t$. We can thus interpret $r_k(t)$ as the fund performance before cash flows. For example, a fund may show a TNA gain of $10\%$ if its portfolio gained $5\%$ and sold $5\%$ of TNA in new sales. Taking $f(t)$ into account solves this bias and retrieves the fund performance.

\begin{equation}
    r_k(t) = \frac{TNA(t) - TNA(t-k)}{TNA(t-k)} - f(t)
\end{equation}

Excess returns are the return of a fund minus the average return of the funds at this period, weighted by their size.\\

Net flows $f(t)$ are the difference between new sales (inflows) and cash redemptions (outflows), divided by the fund TNA.

\begin{equation}
    f(t) = \frac{I(t) - O(t)}{TNA(t)}
\end{equation}

\subsubsection{N-SAR SEC data}

\paragraph{Data} \cite{ha2019misspecifications} evaluated return discrimination between funds and the investor substitution effect in this flow-performance relationship: the tendency of some investors to redeem the worst performing funds to move to better-performing funds. For this task, they examine actual net flows, inflows and outflows using the Form N-SAR filings from the Securities Exchange Commission EDGAR database and returns from the Center for Research in Security Prices (CRSP) mutual fund database. \par

Form N-SAR A/B is a semi-annual report that registered investment companies must file. The SEC EDGAR database started in 1994; the sample period is from January 1994 to June 2015. The data focuses on US domestic equity mutual funds in growth, growth\&income and mid-small-cap fund styles, excluding exchange-traded funds (ETFs) and index funds. The data also excludes funds with less than \$15M in total assets under management (TNA) and aged less than three years. The data after this screening process contains 3562 domestic equity funds.

\paragraph{Results}

We regress monthly net flows as a function of past returns above the benchmark (the weighted average return of funds), using various time lags (1M, 6M, 1Y, 2Y, 3Y, 4Y, 5Y, 10Y). We include some controls: fund age, fund TNA, investment style, fund expense ratio and time (months). Table 3 shows the results.

We observe a negative constant of $-0.0164$ ($p=0.002$), meaning that, on average, funds lose 1.6\% of their TNA regardless of performance. This negative constant can correspond to redemptions because of liquidity needs. We can also interpret it as the ongoing shift from actively managed funds to index funds or bonds. Finally, portfolio rebalancing may affect the constant since equity gains are usually higher than bonds. If equity gains have been substantial, investors following a 60-40 equity-bond allocation end up overweight in equity and must redeem some fund shares to return to their target. Only the 1-month and 10-year lags are significant among all lags, and both positively impact net flows. One percentage point of a 10-year return increases the fund TNA by 0.43\%.

However, R-squared is very low, and the data present excess skewness and kurtosis, which suggests that the above model is not well specified or that the data is very noisy due to high-frequency self-reporting. We move to the aggregate ICI data to confirm our findings on monthly flows.

\begin{table}
\begin{center}
\begin{tabular}{lclc}
\toprule
\textbf{Dep. Variable:}    &    Net\_flows    & \textbf{  R-squared:         } &         0.000      \\
\textbf{Model:}            &       OLS        & \textbf{  Adj. R-squared:    } &         0.000      \\
\textbf{Method:}           &  Least Squares   & \textbf{  F-statistic:       } &         1.056      \\
\textbf{Date:}             & Wed, 14 Sep 2022 & \textbf{  Prob (F-statistic):} &        0.391       \\
\textbf{Time:}             &     14:30:06     & \textbf{  Log-Likelihood:    } &    -1.2203e+05     \\
\textbf{No. Observations:} &       73380      & \textbf{  AIC:               } &     2.441e+05      \\
\textbf{Df Residuals:}     &       73371      & \textbf{  BIC:               } &     2.442e+05      \\
\textbf{Df Model:}         &           8      & \textbf{                     } &                    \\
\textbf{Covariance Type:}  &    nonrobust     & \textbf{                     } &                    \\
\bottomrule
\end{tabular}
\begin{tabular}{lcccccc}
                        & \textbf{coef} & \textbf{std err} & \textbf{t} & \textbf{P$> |$t$|$} & \textbf{[0.025} & \textbf{0.975]}  \\
\midrule
\textbf{const}          &      -0.0164  &        0.005     &    -3.039  &         0.002        &       -0.027    &       -0.006     \\
\textbf{exc\_return1}   &       0.1630  &        0.095     &     1.721  &         0.085        &       -0.023    &        0.349     \\
\textbf{exc\_return6}   &      -0.0006  &        0.019     &    -0.033  &         0.974        &       -0.038    &        0.037     \\
\textbf{exc\_return12}  &      -0.0021  &        0.023     &    -0.093  &         0.926        &       -0.047    &        0.042     \\
\textbf{exc\_return24}  &      -0.0014  &        0.016     &    -0.086  &         0.932        &       -0.033    &        0.030     \\
\textbf{exc\_return36}  &       0.0089  &        0.014     &     0.642  &         0.521        &       -0.018    &        0.036     \\
\textbf{exc\_return48}  &      -0.0042  &        0.012     &    -0.336  &         0.737        &       -0.029    &        0.020     \\
\textbf{exc\_return60}  &      -0.0012  &        0.008     &    -0.157  &         0.875        &       -0.017    &        0.014     \\
\textbf{exc\_return120} &       0.0043  &        0.002     &     1.961  &         0.050        &     2.11e-06    &        0.009     \\
\bottomrule
\end{tabular}
\begin{tabular}{lclc}
\textbf{Omnibus:}       & 302458.360 & \textbf{  Durbin-Watson:     } &         1.717      \\
\textbf{Prob(Omnibus):} &    0.000   & \textbf{  Jarque-Bera (JB):  } & 2852915030912.781  \\
\textbf{Skew:}          &  -111.696  & \textbf{  Prob(JB):          } &          0.00      \\
\textbf{Kurtosis:}      & 30548.664  & \textbf{  Cond. No.          } &          67.6      \\
\bottomrule
\end{tabular}
\caption{OLS Regression Results of monthly net flows with SEC-N-SAR form data}
\end{center}
\end{table}

\subsubsection{ICI data}

\paragraph{Data} Another way of estimating the fund flow relationship is to use aggregated data \cite{ici22}. Table 17 of the 2022 ICI FactBook, "Long-Term Mutual Funds: Net New Cash Flow, Total Net Assets, and Flows as a Percentage of Previous Year’s Total Net Assets", allows to aggregate flows as a percentage of previous year TNA and regress those on the TNA change, from 1995 to 2021, distinguishing between equity, hybrid and bond funds. Unlike our last results, the data contain annual net flows.

\paragraph{Results} Using this data and computing annual lagged returns at 2, 5 and 10 years horizons,  the 10-year returns significantly positively impact net flows, as shown in Table 4. One percent return at a 10-year horizon increases fund TNA by 53\% ($p=0.06$). Other lags do not appear significant. The constant is significant ($p=0.015$) and negative with a high magnitude: mutual funds annually appear to lose more than 2.4\% of their TNA regardless of performance. This negative constant can correspond in the data to the long-term outflows from actively managed mutual funds to index funds, investor portfolio rebalancing and liquidity shocks.

\begin{table}
\begin{center}
\begin{tabular}{lclc}
\toprule
\textbf{Dep. Variable:}    &    Net Flows     & \textbf{  R-squared:         } &     0.665   \\
\textbf{Model:}            &       OLS        & \textbf{  Adj. R-squared:    } &     0.601   \\
\textbf{Method:}           &  Least Squares   & \textbf{  F-statistic:       } &     10.42   \\
\textbf{Date:}             & Wed, 14 Sep 2022 & \textbf{  Prob (F-statistic):} &  8.23e-05   \\
\textbf{Time:}             &     14:19:07     & \textbf{  Log-Likelihood:    } &   -66.244   \\
\textbf{No. Observations:} &          26      & \textbf{  AIC:               } &     142.5   \\
\textbf{Df Residuals:}     &          21      & \textbf{  BIC:               } &     148.8   \\
\textbf{Df Model:}         &           4      & \textbf{                     } &             \\
\textbf{Covariance Type:}  &    nonrobust     & \textbf{                     } &             \\
\bottomrule
\end{tabular}
\begin{tabular}{lcccccc}
                     & \textbf{coef} & \textbf{std err} & \textbf{t} & \textbf{P$> |$t$|$} & \textbf{[0.025} & \textbf{0.975]}  \\
\midrule
\textbf{const}       &      -2.4610  &        0.933     &    -2.638  &         0.015        &       -4.401    &       -0.521     \\
\textbf{return\_1Y}  &       0.0175  &        0.043     &     0.403  &         0.691        &       -0.073    &        0.107     \\
\textbf{return\_2Y}  &       0.0420  &        0.031     &     1.361  &         0.188        &       -0.022    &        0.106     \\
\textbf{return\_5Y}  &       0.0080  &        0.014     &     0.577  &         0.570        &       -0.021    &        0.037     \\
\textbf{return\_10Y} &       0.0053  &        0.003     &     1.961  &         0.063        &       -0.000    &        0.011     \\
\bottomrule
\end{tabular}
\begin{tabular}{lclc}
\textbf{Omnibus:}       &  1.195 & \textbf{  Durbin-Watson:     } &    0.686  \\
\textbf{Prob(Omnibus):} &  0.550 & \textbf{  Jarque-Bera (JB):  } &    0.323  \\
\textbf{Skew:}          &  0.202 & \textbf{  Prob(JB):          } &    0.851  \\
\textbf{Kurtosis:}      &  3.367 & \textbf{  Cond. No.          } &     907.  \\
\bottomrule
\end{tabular}
\caption{OLS Regression Results of annual flows using ICI data}
\end{center}
\end{table}

\subsection{Supplementary details on the benchmark learning tasks}

\subsection{Interpretability and robustness}

Although interpretability in this context requires further definition, we can start by observing that real-world strategies usually rest on some economic reasoning. For example, trend followers assume that price trends are persistent. Interpretable strategies should exhibit a generalisable ``model'' of the financial market. As for robustness, the strategies can readily train on various market initial conditions or face unseen policy interventions and structural model changes, which are easy to implement in the ABM. For example, in the current context of rising interest rates, we could train a strategy on a probability distribution of interest rates rather than on a single interest rate value. Finally, the Evology environment would be more useful if its insights could generate useful signals for trading in real markets, an additional challenge to robustness and ABM realism. Since the Lucas critique, we know that changes in economic policies can lead to structural changes in modelling and optimal behaviour. Since we are in a simulation environment, we can alter market conditions, and training can include those changes. It is easy to implement changes in the interest rates, dividend policies and other interventions in an ABM. For example, Central Banks' quantitative easing can involve a fictitious agent with a constant positive excess demand for the asset.

\subsection{Task 1: trading strategy}

The trading is simplified to arbitrage between the asset and cash. Given a set of market conditions $M$, the action space available to our adaptive fund is a $T$-dimensional vector $\Phi$ of trading signal values. On each day $t$, the fund decides if it is buying or selling asset shares. 

\begin{equation}
    \Phi = [\phi(0), \phi(1), \dots, \phi(T)], \phi(t) \in [-1, 1] \ \forall t
\end{equation}

The optimisation problem faced by the adaptive fund consists in finding the sequence of trading signals $\Phi$ that maximises some measure of performance $\mathcal{P}$, typically a measure of profits such as the Sharpe ratio (geometric mean of returns divided by the standard deviation of returns) or wealth multipliers (ratio of increase of the funds' wealth after $T$ simulation steps). 

\begin{equation}
    \max_\Phi \mathcal{P}
\end{equation}

The baselines to beat are first the performance of the base strategies introduced in the model. The first definition of success for the adaptive fund optimising a trading strategy is to achieve better performance than the NT, VI and TF strategies present in the model, under the same bounded rationality limitations. This environment requires the agent to understand its market impact and the returns landscape and to learn to mitigate the effect of its size on its returns. Once we achieve this target, setting a benchmark for simulation performance is more difficult as this benchmark is new. However, obtaining Sharpe ratios superior to empirical hedge fund levels, i.e. 10 or 20, would be a good starting point. \par 
There are several interesting extensions of this task. The first is to evaluate the performance of different optimisation goals $\mathcal{P}$. Does profit maximisation lead to the highest profits, or is the problem so deceptive that other metrics, multi-objective fitness, or even novelty search \cite{lehman2011novelty} can lead to better strategies? We can also consider two sub-tasks for Task 1. The optimised trading strategy could be a single trading signal function $\phi(t)$ kept fixed during each evaluation: this would be a \textit{static} task. In a more \textit{dynamic} approach, the agent could instead evolve a sequence of different trading signal functions. Learning thus would focus on the meta-strategy that governs the adoption at any period of a specific trading signal.

\paragraph{Task 2: investment strategy optimisation}

The investor is learning what characteristics of the funds should determine their inflows and outflows so that the investor achieves the highest return on their investment. Using a similar choice of the performance measure $\mathcal{P}$, the investor maximisation problem is:

\begin{equation}
    \max_\nu \mathcal{P}
\end{equation}

Where $\nu$ is a function that maps an investment decision (the quantity of money to invest in or out of a fund) based on some fund characteristics $\textbf{x}$, including measures of fund return at various timescales, the statistical significance of those returns, fund size, fund strategy... It will be fascinating to consider how this evolved investor behaves concerning timescales \cite{scholl2021market}, compare its performance to typical investment firms and how statistically significant their actions are. The first baseline will be to achieve a higher success (e.g. profitability or rationality) than the average investor behaviour implemented in the model and derived from empirical data \cite{ha2019misspecifications}.

\subsection{Supplementary conclusions and limitations}

\paragraph{Modelling approach} 
Modelling implies identifying the critical characteristics of the industry to obtain a model that is simple enough to allow analysis but realistic enough to be interesting. Two main approaches are possible to model the ecology of mutual funds. The first is to focus on aggregate quantities of each investment style and deal with a few agents, each representing Small-cap Value, Large-cap Growth, and so on. This representation is possible with the total net assets (TNA) data provided by \cite{ici22}. For instance, the 168 small-cap growth mutual funds reduce to a single representative agent with a TNA of 315,759 million dollars. A second approach would be to model each of those 168 funds and do the same for all investment styles. This second approach, more realistic, is much more data intensive. It is likely that no pair of funds in the same style have the same strategy, expense ratios, and performance history. Morningstar, Lipper, SEC's EDGAR and the Center for Research on Securities Prices (CRSP) can provide such detailed fund-level data at some cost. In particular, this second approach needs fund-level data on the portfolio strategy. If the data does not distinguish those strategies, those 168 identical agents are equivalent to a single agent aggregating the 168 funds TNAs. For data availability reasons, we are thus currently taking the first approach. We consider the diversity of equity investment styles covered by the ICI data and use representative, stylised agents for the different investment styles: growth, value and blend. One additional common distinction is also the market capitalisation of the stocks in the funds' portfolio: small-cap (company market capitalisation between 300 million and 2 billion), mid-cap (2-10 billion), large-cap (>10 billion, e.g. Apple) and multi-cap.

\paragraph{Possible developments of the model to other participants}
Table \ref{facts} focuses on actively-managed Equity mutual funds. The three large categories in the ICI classification are Equity, Hybrid and Bonds. Equity (domestic and the world) represented more than 14.7 trillion dollars, hybrid 1.8 and bonds 5.5. Money-market funds, which invest in high-quality, short-term debt, represent 4.7 trillion dollars. It is worth noting that the current model does not consider the bond markets. 
Other participants may also reveal themselves as necessary to include in the model. In particular, index funds (index mutual funds and index ETFs) have recently gained in size and market share, notably for their lower expense ratios. In the US alone, ETFs represent 7.2 trillion dollars in market share. Actively-managed mutual funds have been experiencing continuous outflows while index funds have received inflows of equal size. Index funds likely impact more and more the interactions between mutual funds. Active funds represented in 2021 year-end more than 16 trillion US\$ (6680 entities) and index funds 5.7 trillion (496 entities). \cite{ici22}.

\paragraph{Improving the mutual fund and securities' population in the ABM}
We outlined in the Introduction some main characteristics of the mutual funds' population and acknowledge that the current model partially meets those characteristics. We identify key steps for our future work to improve validation in this aspect:
\begin{itemize}
    \item Model entry and exit of funds (in 2021, 748 funds opened and 457 were merged or liquidated),
    \item Fit the model investment styles to the data with Value, Growth and Blend,
    \item Multiply and diversify model assets to obtain small, mid and large-cap stocks attached to a realistic set of companies at various stages of their expansion. We would like a diverse set of companies that feature value and growth,
    \item Ensure that the distribution of fund returns in the model is consistent with empirical returns at various time horizons,
    \item Adjust the initial conditions, i.e. the market composition in terms of TNA, to the ICI data,
    \item Consider the necessity of funds dividends, reinvestment and expense ratios,
    \item Complement the ICI data with other classification methods to include more diverse strategies, particularly ETFs, index funds and the bond market.
\end{itemize}

\paragraph{Next steps} A potential goal for this simulation could be to become an easy-to-use toolkit for developing machine learning algorithms, just like the Open AI gym \cite{brockman2016openai}. It is thus essential that the environment offers sufficient complexity to be of interest to the machine learning community and is realistic enough for its insights to be potentially transferable to the real world. The simulation should be efficient enough to allow training, e.g. through developing a GPU implementation. Calibration and validation efforts should continue to emulate market environments, mainly populating the ecology with realistic strategies reconstructed from actual portfolios \cite{scholl2021market}. Thanks to publicly available SEC data, model calibration constitutes a benchmark task of interest. One task of potential interest that is not mentioned yet in the model is the optimisation of financial regulation: we could easily add to the model described in Figure 2 an additional regulation component in which a policy-maker observes the market and attempts to limit volatility, inflation or other measures of concern, by restricting agent behaviour through maximum leverage, price interventions such as quantitative easing or tightening. This addition would open a new task in which a machine learning algorithm can attempt to optimise financial regulation to achieve those goals.

\paragraph{Next-level calibration with simulation-based inference} Although we have reduced the number of free parameters in the model, we are left with several free parameters of importance. We assumed that the various heterogeneous sub-strategies (e.g. time horizon $\theta_i$, required rates of return $\Tilde{r}_i$) are uniform within arbitrary ranges. The leverage of the various strategies is also significant to the ecology dynamics as they impact the system's position in the profit landscape. Several of the abovementioned updates may also bring more free parameters to tackle. A poor choice of summary statistics can lead to losing information from the empirical data and reduce the quality of calibration \cite{dyer2022calibrating}, which motivates alternatives such as simulation-based inference using graph neural networks. These procedures will likely require further exploration of empirical data, such as the N-PORT forms and 13F filings.

\end{document}